# RSU Cloud and its Resource Management in support of Enhanced Vehicular Applications

Mohammad A. Salahuddin, Ala Al-Fuqaha, †Mohsen Guizani and ‡Soumaya Cherkaoui
Department of Computer Science, Western Michigan University, Kalamazoo, MI, USA
†Qatar University, Doha, Qatar
‡Université Sherbrooke, Sherbrooke, QC, Canada
{mohammad.salahuddin, ala.al-fuqaha}@wmich.edu

*Abstract*—We propose Roadside Unit (RSU) Clouds as a novel way to offer non-safety application with QoS for VANETs. The architecture of RSU Clouds is delineated, and consists of traditional RSUs and specialized micro-datacenters and virtual machines (VMs) using Software Defined Networking (SDN). SDN offers the flexibility to migrate or replicate virtual services and reconfigure the data forwarding rules dynamically. However, frequent changes to service hosts and data flows not only result in degradation of services, but are also costly for service providers.

In this paper, we use Mininet to analyze and formally quantify the reconfiguration overhead. Our unique RSU Cloud Resource Management (CRM) model jointly minimizes reconfiguration overhead, cost of service deployment and infrastructure routing delay. To the best of our knowledge, we are the first to utilize this approach. We compare the performance of purist approach to our Integer Linear Programming (ILP) model and our innovative heuristic for the CRM technique and discuss the results. We will show the benefits of a holistic approach in Cloud Resource Management with SDN.

## I. INTRODUCTION

Intelligent Transportation Systems (ITS) comprise services and applications for Vehicular Ad hoc Networks (VANETs) to increase the safety and efficiency of our road infrastructure. VANET nodes consist of On-board Units (OBUs) in vehicles and fixed Roadside Units (RSUs) in infrastructure. While OBUs comprise computational, communicational, storage, sensory and positioning systems, RSUs can vary in size and form, ranging from small, resource-constrained roadside mounted traffic monitoring cameras to high-power communication towers. Wireless Access in Vehicular Environment (WAVE) standards and protocols dictate how vehicles communicate with other vehicles (V2V) and infrastructure (V2I). Generally, time critical safety applications use V2V, while non-safety applications rely on V2I [1].

We propose an innovation to the traditional VANET infrastructure, with RSU Clouds. This Cloud architecture consists of traditional RSUs and specialized micro-datacenters and virtual machines (VM) that leverages the deep programmability of Software Defined Networking (SDN). Cloud service providers offer infrastructure and applications that can be *rented* for hosting applications and services. The network can be *reconfigured*, at a cost ( [2],[3],[4] ), so that the virtual services can be migrated or replicated to meet *fluctuating* demands or efficiently use Cloud resources by exploiting the flexibility of SDN. In SDN, we can program controllers, in the control plane, to define the data forwarding rules for the switches in the data plane.

Due to the fixed infrastructure, SDN and proximity to end-users [5], RSU Clouds offer a reliable and flexible way to meet the QoS required by the fluctuating demands. However, it is a non-trivial problem that requires multiple design considerations. First, it requires a cost effective solution for hosting, migrating and replicating services in the network. Second, RSU Clouds must be able to maintain the QoS for applications. Third, they must employ an efficient reconfiguration scheme to adapt to the frequent change in service demands. We define a novel and efficient RSU Cloud Resource Management (CRM) model that minimizes the reconfiguration overhead, while minimizing the number of service replications and infrastructure delay and meeting network and link layer bounds.

Therefore, the scope of this work and its contributions are:
- Architecture definition for RSU Clouds and its micro-datacenters;
- Reconfiguration overhead analysis by emulating an OpenFlow enabled SDN in Mininet;
- Definition of the CRM model and its resolution as a multi-objective Integer Linear Programming (ILP) problem;
- Design of an efficient heuristic for CRM;
- Performance comparison of a purist approach with CRM.

The rest of this paper is organized as follows. Section II reviews and compares RSU Clouds with related work. Section III delineates the details of the RSU Cloud architecture and its micro-datacenters and reconfiguration overhead analysis. Section IV presents the CRM model and Section V highlights the results and contributions of CRM and RSU Clouds. We conclude with an overview of our contributions in Section 0.

## II. RELATED WORK

A brief overview of the related work in the areas of Cloud Computing in VANETs and Cloud Resource Management will highlight the novelty of our RSU Cloud architecture and its RSU Cloud Resource Management scheme.

## A. Cloud Computing in VANETs

Integrating VANETs with the Cloud increases utilization of myriad computational capabilities that are underutilized by safety applications alone ( [6], [7] ) and overcomes the routing issues in V2V communications [8]. Lee *et al.* [9] aim to interconnect OBU and RSU resources into a Cloud for cooperative sensory, storage and computing tasks, while others in [6] and [10] propose that Road-Side Units (RSUs) act as gateways to traditional Clouds or design a Cloud of On-board Units (OBUs), respectively. Vehicular Cloud Networking (VCN) [9] is being proposed as a revolution to modernize the traditional VANET, which integrates information centric networking and Cloud Computing with VANETs. In VCN, vehicles and resource constrained RSUs share their resources in one virtual platform.

Our proposed RSU Cloud architecture and its SDN differs from these in two-fold, (1) it consists of traditional RSUs and very small scale datacenters, and (2) it can be dynamically reconfigured, to cost effectively meet changes in service demands while guaranteeing QoS. Furthermore, our RSU Clouds can be easily adapted to any VANET Cloud paradigm.

## B. Cloud Resource Management

The current techniques for efficient and effective management of Cloud services include rich connectivity at the edge of the network and dynamic routing protocols that reduce routing delay and balance traffic [11]. Presently, resource management techniques are employed via capacity planning tools such as VMWare Capacity Planer™, IBM Websphere Cloudburst™, Novell PlateSpin Recon™, etc. that decide the location of VM placement [11]. However, they lack in load balancing at VM and result in highly imbalanced traffic distribution [11]. Various researchers ( [12], [13], [14] ) proposed solutions for deployment of Cloud services for low latency, either independently or jointly with cost of deployment.

On the other hand, others ( [2], [4], [15], [16] ) have studied the cost of VM migration and reconfiguration with SDN. VM migrations can deteriorate network performance [2], while modification to the data forwarding rules yield traffic congestions at the controller in the control plane [4].

In contrast to the Cloud resource management techniques mentioned, we optimize the network reconfiguration overhead, *while* jointly minimizing the cost of service deployment and infrastructure routing delay. To the best of our knowledge, we are the first to jointly minimize reconfiguration overhead, cost of service deployment and infrastructure routing delay.

## III. RSU CLOUD

In this section, we will present our RSU Cloud architecture with OpenFlow, the defacto standard protocol for SDN. In SDN, controllers in the control plane define forwarding rules for switches in the data plane. Each switch receives flow rules, proactively or reactively, from controllers, via the control plane. This separation of data and control plane enables SDNs to be dynamically reconfigured [3].

### A. Architecture

Users can increase in-vehicle productivity by subscribing to convenience and infotainment services such as remote vehicle diagnostics, on-the-go-Internet, online gaming, multimedia streaming and voice-over-IP. Non-safety applications and services can attract significant attention and thrust ITS forward [17]. However, although there is no stringent QoS requirement for non-safety applications, it is highly desirable and often crucial for its users [17]. Therefore, we design RSU Clouds that infringe upon the fundamentals of Fog Computing and push the services to the edge of the network, to increase reliability and QoS, with respect to latency.

Our RSU Cloud, illustrated in Fig. 1(a) includes traditional RSUs and micro-datacenters that will host the services to meet the demand from the underlying OBUs in the mobile vehicles. Our RSU micro-datacenter, illustrated in Fig. 1(b), is a traditional RSU with additional components for virtualization and SDN. The micro-datacenter hardware consists of a small form factor computing device and an OpenFlow switch. The software components on the computing device include the host operating system and a lightweight hypervisor. A hypervisor is a low-level middleware that enables virtualization [2], so that VMs can efficiently share resources to host services that can be migrated and, or replicated.

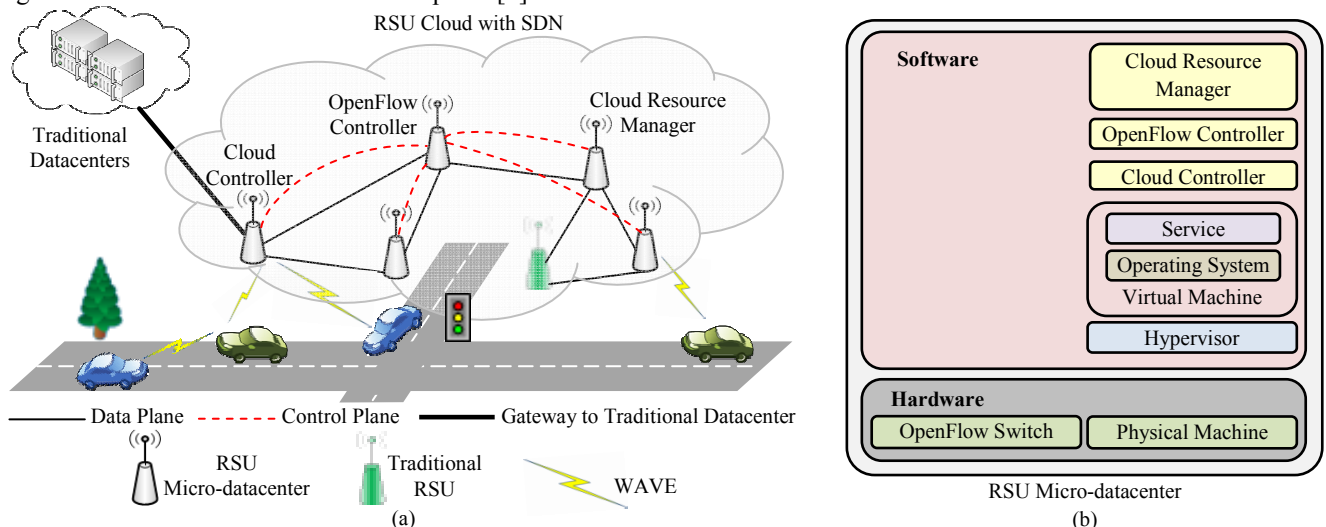

Fig. 1. Architecture of (a) RSU Cloud and (b) RSU Micro-datacenter.

Some of the micro-datacenters will have additional software components, like OpenFlow Controller(s), Cloud Controller(s) and RSU Cloud Resource Manager(s). Our novel RSU Cloud Resource Manager (CRM) communicates with OpenFlow and Cloud controllers, via the data plane, to disseminate information regarding service hosting, service migration and, or data flow changes, as illustrated in Fig. 1(a). In the data plane, Cloud Controllers will govern service migration and hypervisors to instantiate or eliminate new VMs hosting services. Consequentially, OpenFlow Controllers will update switch flow rules via the control plane.

The dynamic service demands from vehicles may require increasing or decreasing the number of micro-datacenters hosting the services or physically migrating the VMs hosting the services from one micro-datacenter to another via the data plane. Without loss of generality, we interchangeably use VM migration and service migration. Though, we can reprogram the RSU Clouds to dynamically update service hosting and data forwarding information, it is costly and deteriorates the network performance ( [2], [3] ). VM migration increases latency in the data plane and modifying data forwarding rules that increases the control plane overhead [4].

Service providers incur the cost of service migration and, or replications, and service users experience the deterioration in QoS. Naïve approaches to hosting services across all micro-datacenters are too costly, since service providers rent cloud resources from cloud infrastructure providers. Our major contribution is the novel offline Cloud Resource Manager, which is responsible for making the optimal decisions regarding the location and number of services hosted and data forwarding rules, in the RSU Cloud over time.

### B. Reconfiguration Overhead Analysis in Mininet

We investigate the reconfiguration overhead by emulating a RSU Cloud with SDN in Mininet [18]. We design a RSU Cloud topology that is inspired by Florida Department of Transportation (FDOT) deployment of RSUs [19]. We begin by implementing a RSU micro-datacenter as a VM host connected to an OpenFlow switch, with a zero delay. The switches maintain data forwarding rules in flow tables. We want to study the cost of reconfiguring the flow tables over time, as the demand changes. We define a configuration as a snapshot of all service hosts and all data forwarding rules.

Evidently, real-world networking scenarios must support multipath. OpenFlow supports multipath with a group table feature, which allows the Controller to specify multiple paths for an incoming packet. The Open vSwitch (OvS) 2.1 supports group tables but implements random path selection. We implemented stochastic switching for multipath support in OpenFlow enabled SDN [20].

Our reconfiguration overhead heuristic counts the changes in the data forwarding rules and the number of service migrations. The control plane overhead includes cost associated with adding, deleting and, or modifying flow or group table rules. Note that a flow table and a group table rule modification is counted doubly, a deletion of the old rule, followed by an addition of a new rule. The VM migration overhead is simplified to counting changes in service hosts, and does not count VM tear down or deleting hosts, since it does not induce network traffic. We can extend the count to be a more complex measure with sensitivity to different parameters such as size of the VMs, time of day, etc.

Based on emulation and reconfiguration overhead analysis, we can formally define reconfiguration overhead. Given, a configuration is a network snapshot that records the service hosts and the data forwarding rules, which consists of flow and group table rules. A configuration at time $t_i$ can be defined as a tuple $< X^{t_i}, Y^{t_i}, Z^{t_i} >$, where $X^{t_i} = \{x_1, x_2, \dots, x_{|X^{t_i}|}\}$ is the set of service hosts, $Y^{t_i} = \{y_1, y_2, \dots, y_{|Y^{t_i}|}\}$ is the set of flow rules and $Z^{t_i} = \{z_1, z_2, \dots, z_{|Z^{t_i}|}\}$ is the set of group rules.

Therefore, given sets of service hosts, $X^{t_i}$ and $X^{t_{i+1}}$ for time $t_i$ and $t_{i+1}$, respectively, VM migrations is in Equation (1). And given sets of flow and group rules $(Y^{t_i}, Z^{t_i})$ and $(Y^{t_{i+1}}, Z^{t_{i+1}})$, for time $t_i$ and $t_{i+1}$, respectively, the control plane overhead is in Equation (2), and is defined as the sum of flow and group rules to be deleted and added, flow rules changed to group rules and group rules changed to flow rules.

$$VM\ migrations = |X^{t_i} - X^{t_{i+1}}| \quad (1)$$

$$\begin{aligned} control\ plane\ overhead = \\ |Y^{t_i} - Y^{t_{i+1}}| + |Y^{t_{i+1}} - Y^{t_i}| + |Z^{t_i} - Z^{t_{i+1}}| + \\ |Z^{t_{i+1}} - Z^{t_i}| + |Y^{t_i} \cap Z^{t_{i+1}}| + |Z^{t_i} \cap Y^{t_{i+1}}| \end{aligned} \quad (2)$$

### IV. CLOUD RESOURCE MANAGEMENT MODEL

#### A. Problem Statement

Given a network graph $G = (V, E)$, set of services $S$, set of average demands $D = \{d_{t_1}, d_{t_2}, \dots d_{t_{|T|}}\}$ over time period $T = \{t_1, t_2, \dots t_{|T|}\}$ with an initial configuration $\psi^{t_1} = \langle X^{t_1}, Y^{t_1}, Z^{t_1} \rangle$ for demand $d_{t_1}$ at time $t_1$. The set $V$ represents the RSU micro-datacenters interconnected by the edges in $E$, each with bandwidth capacity $C_e \forall e \in E$. At time $t_i \in T$, there is a demand $b_{n,k}$ for service $k$, $k \in S$ at node $n$, $n \in V$ and the average demand in the network is $d_{t_i}$. The problem is to find a Pareto Optimal configuration from a set of Pareto Optimal Frontier (POF) of configurations $\Psi^{t_i}$ to minimize the number of VM migrations $\langle \psi_j^{t_i} | \min(X_j^{t_i} - X^{t_{i-1}}), \forall \psi_j^{t_i} = < X_j^{t_i}, Y_j^{t_i}, Z_j^{t_i} >, \psi_j^{t_i} \in \Psi^{t_i} \rangle$. Each $\psi_j^{t_i} \in \Psi^{t_i}$, optimally hosts services within threshold $\phi_k, \forall k \in S$, to meet service demands at $t_i$, while achieving a load-balanced network, and minimizing infrastructure delay with QoS bounds $\sigma_k, \forall k \in S$.

#### B. Delay

The delay model is a Lookup Table (LUT) with interval $\varphi$, which controls the granularity. The granularity of the LUT is a tradeoff to performance. In practice the LUT table will be built over time, from experimental data. However, without loss of generality and similar to ( [21], [22] ), we currently use a G/G/1 queuing system for computing the delay on an edge, as the sum of processing delay which is considered fixed and having a certain value (ex. 10μs), queuing delay ($T_q$), and transmission and propagation delay depending on packet size (ex. packet size 800B) which can also be considered fixed. For modeling this, we assume a Poisson process for packet inter-arrival times $\lambda$ and processing times $\mu$, with mean and standard deviation, $t_a, \sigma_a$ and $t_s, \sigma_s,$ respectively. Our LUT accounts for varying inter-arrival and packet processing times by employing high coefficients of variation in inter-arrival and packet processing times, $c_a = \sigma_a/t_a$ and $c_s = \sigma_s/t_s$, respectively. We use Kingsman formula [23] to approximate the queuing delay, in Equation (3).

$$T_q = \frac{c_a^2 + c_s^2}{2} \cdot \frac{\lambda/\mu}{\mu - \lambda} \quad (3)$$

*C. Formulation*

We leverage multi-objective Integer Linear Programming (ILP) to formulate the optimizations required by RSU Cloud resource management problem. The novelty of our ILP model lies in jointly minimizing the reconfiguration overhead, pertaining to VM migrations, control plane modifications, number of service hosts and cloud infrastructure delays. The objective for minimizing reconfiguration overhead in (4) is the minimization of the number of VM migrations and control plane modifications. We use weight $\rho$ to control the priority of reconfiguration overhead, such that, minimizing VM migrations takes priority over control plane modifications.

$$\min \left\{ \begin{array}{l} \sum_{m=1}^{|V|} \sum_{k=1}^{S} \rho \cdot \alpha_{m,k} + \\ \sum_{m=1}^{|V|} \sum_{n=1}^{|V|} \sum_{x=1}^{k_{m,n}} \sum_{k=1}^{S} (1-\rho) \cdot (\beta_k^{m,n,x} + \gamma_k^{m,n,x}) \end{array} \right\} \quad (4)$$

Where, $\alpha_{m,k} \; \forall \; 1 \leq m \leq |V|, 1 \leq k \leq S$ is a binary indicator of change in service hosts (c.f. (1)). Therefore, sum of $\alpha_{m,k}$ is the overhead due to VM migrations. The addition and deletion of data forwarding rules are in binary indicators $\beta_k^{m,n,x}$ and $\gamma_k^{m,n,x} \; \forall 1 \leq m, n, m \neq n \leq |V|, 1 \leq k \leq S, 1 \leq x \leq k_{m,n}$, respectively (c.f. (2)). Therefore, sum of $\beta_k^{m,n,x}$ and $\gamma_k^{m,n,x}$ constitute the control plane overhead.

Jointly, we achieve a load balanced network and minimize the number of service hosts and infrastructure delay in (5).

$$\min \left\{ \sum_{m=1}^{|V|} \sum_{k=1}^{S} \omega \cdot h_{m,k} + \sum_{e=0}^{|E|} (1-\omega) \cdot \frac{d_e}{q_{C_e,e}} \right\} \quad (5)$$

Where, sum of $h_{m,k}, \forall 1 \leq m \leq |V|, 1 \leq k \leq S$ is the number of service hosts and sum of $d_e, \forall 1 \leq e \leq |E|$ is the total delay on the edges in the network. We can jointly consider two disparate quantity functions like, *delay* on an edge and the *number* of service hosts, if we normalize the delay by the maximum delay on an edge, making it unit-less. Here, we use the weighted sum approach with weight $\omega$. Currently, we count the service hosts, but this can be extended to include a more detailed cost of hosting services with respect to the hardware and software requirements of the services.

Due to the inherent nature of the problem, we decouple the objective in (4) and (5). We use lp_solve to optimize (5), subject to the following constraints (i) all service requests must be satisfied; (ii) the load on an edge in the network is bounded by bandwidth capacity of the edge; (iii) the load on the edge must be a multiple of $\varphi$, the LUT interval; (iv) the service host can serve requests locally with zero infrastructure delay; (v) QoS bounds $\sigma_k$; (vi) number of service host bounds $\phi_k$. We solve this ILP and populate a POF of non-dominated solutions, that is, configurations. For our objective in (4), we select a configuration from the POF, such that, it minimizes VM migrations followed by control plane overhead.

*D. Heuristic*

We design and implement a novel heuristic for the RSU cloud resource management problem. We will show that our heuristic efficiently yields results, by *always* operating on the Pareto Optimal Frontier of non-dominated solutions. Our heuristic can be decomposed into two components, (1) generate POF, and (2) prune POF to find the configuration that minimizes VM migrations, followed by the control plane overhead. It is important to note that for a given average demand $d_{t_i}$, we use the *same* demand ($b_{n,k}$) for each RSU as in the ILP. This ensures accurate comparison of the results from ILP optimization and our proposed heuristic.

To generate POF, we begin by randomly selecting $\lceil |V|/2 \rceil$ nodes to be the RSUs hosting services, for each service $k \in S$. The hosts can meet their demands locally. For all remaining RSUs requesting service, we randomly select a RSU host $n \in V$ to satisfy a unit of demand requested. This is repeated *iteratively* so that every unit of demand receives the best infrastructure delay. This uniquely enables us to distribute the load across the paths in the networks and achieve a load-balanced network. We repeat this, until all the demands from all the RSUs have been satisfied. This yields a configuration, for the service hosts and the required data forwarding rules. We repeat this, to generate K configurations and select the configuration that minimizes the infrastructure delay, to be included in the POF.

Next, we reduce the number of service hosts by half, and repeat the process until the POF, $\Psi^{t_i}$, has been filled with all the valid configurations for an average demand $d_{t_i}$ at $t_i$. The POF generation takes $O(Klog|V|)$, making it scalable with the number of nodes in the network. From the POF, we select the configuration that minimizes the number of VM migrations, followed by control plane modifications.

## V. RESULTS

In this section, we will present and discuss our results. We setup the topology inspired from FDOT RSU deployment [19] and use it for comparing the performance of RSU Cloud resource management with purist cost approach. In our scenario, $|V| = 10$, $S = 1$, for a time period T, the set of average demands $D = \{50Mbps, 60Mbps, 80Mbps, 70Mbps, 90Mbps, 50Mbps, 70Mbps\}$, with standard deviation $\sigma = 0.05$, Fast Ethernet connections so bandwidth capacity of each edge $C_e = 100Mbps, \forall 1 \leq e \leq |E|$ and the LUT interval $\varphi = 1$, so that we have a fine grain LUT.

We compare our results with a purist approach. In the given problem, there are two possible purist approaches (1) Cost optimization, which optimally hosts services to meet network demands, irrespective of the infrastructure delay incurred by the services, and (2) Delay optimization, which optimally hosts services to minimize infrastructure delay, irrespective of the cost of hosting services. Trivially, Delay optimization would maximally deploy hosts, so that services are met locally, incurring no infrastructure delay. Moreover, this would not be viable for RSU Cloud service providers. Therefore, we do not include this in our comparison.

Fig. 2 depicts the significant reduction in the number of VM migrations with CRM heuristic with K=100 and CRM optimization. Note, though the CRM heuristic works on the

Pareto Frontier, it may be working on a different configuration when compared to CRM optimization and purist approach.

However, Fig. 3, illustrates that a Heuristic with K=100, performs the worst with respect to the control plane overhead. This is because of our load-balanced CRM heuristic that distributes single units of load across paths until the demands are met. Each path accounts for numerous control plane modifications, as evident by the high cost of control plane modifications. We increase utilization of paths, reduce bottlenecks and potential starvation of other RSUs, in contrast to the naïve approach of selecting the shortest path and fully utilizing the capacity of a path(s).

The benefits of our CRM Optimization are highlighted in Fig. 4 and Fig. 5. Fig. 4 illustrates how closely CRM optimization and CRM heuristic with K=100 perform when compared to the purist cost optimization for hosting services. In Fig. 5, CRM heuristic outperforms optimization with respect to delay, because the best configuration selected had more service hosts (more costly deployment), as in Fig. 4.

Fig. 6 illustrates that each $\psi_j^{t_i} \in \Psi^{t_i}$ is a Pareto Optimal configuration that minimizes infrastructure delay and number of service hosts. Essentially, the heuristic is selecting the configuration that minimizes VM migrations and is always operating on the Pareto Optimal Frontier. Therefore, the final configuration selected *is* also an optimal configuration of the number of service hosts and infrastructure delay.

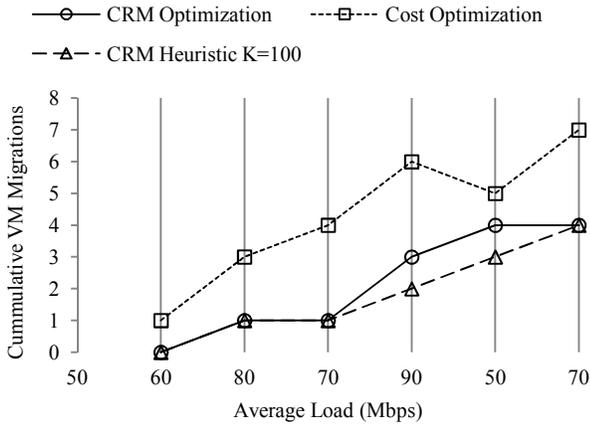

Fig. 2. CRM optimization and heuristic with K=100 outperform purist.

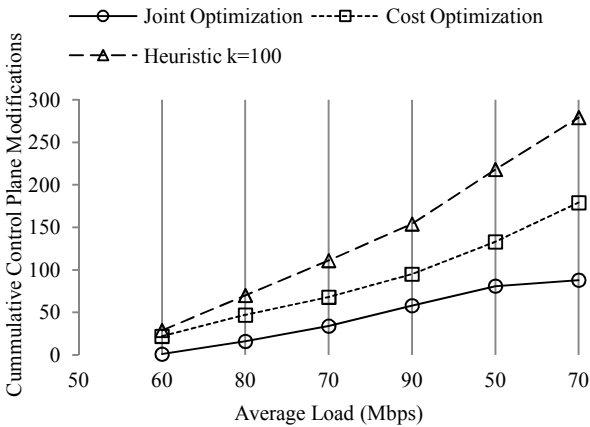

Fig. 3. Effect of fine grain load balancing in CRM heuristic.

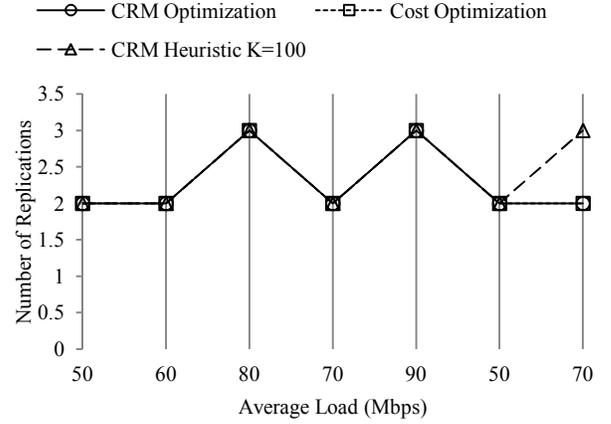

Fig. 4. Purist cost and CRM optimization perform better than CRM heuristic.

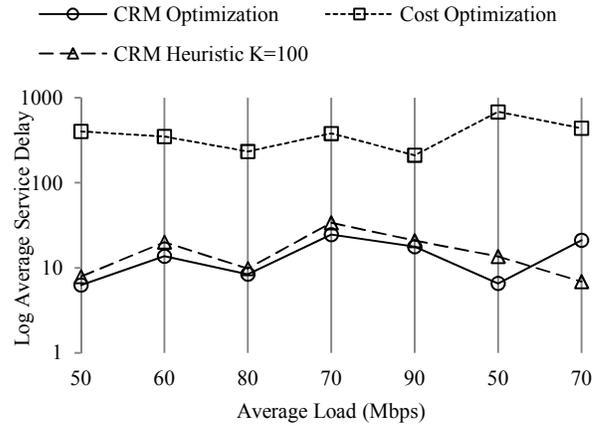

Fig. 5. Efficient CRM heuristic yields suboptimal infrastructure delay.

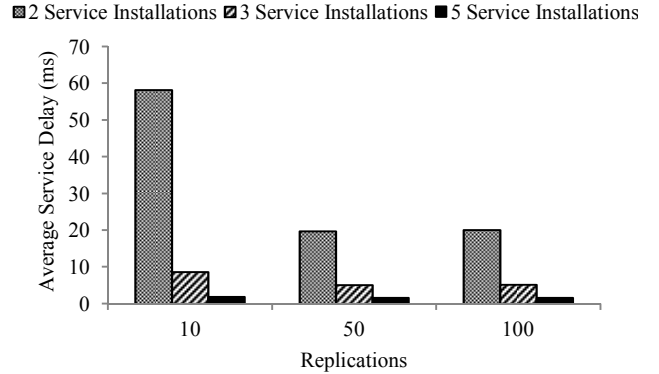

Fig. 6. Every $\psi_j^{t_i} \in \Psi^{t_i}$ is a Pareto Optimal configuration with respect to number of service hosts and infrastructure delay.

Fig. 7 and Fig. 8 illustrates that as the number of replications (K) increases, results show that the configurations are improving in minimizing VM Migrations and control plane overhead. Therefore, we use the CRM heuristic with K=100, as the best results to compare with CRM optimization and purist cost optimization.

## VI. CONCLUSION

In this research, we propose a RSU Cloud to guarantee QoS for non-safety ITS applications. RSU Clouds consist of traditional RSUs and specialized RSUs containing micro-datacenters with SDN. RSU Clouds can be reconfigured, at a cost, to meet the fluctuating service demands. We study the effects of reconfiguration on the SDN and formally define reconfiguration overhead, in terms of VM migrations and control plane modifications. Frequent reconfigurations deteriorate network performance and QoS of ITS applications and services. We define an efficient RSU Cloud resource management technique that jointly minimizes VM migrations, control plane overhead, number of service hosts and infrastructure delay. To the best of our knowledge, we are the first to design such a model.

We illustrate how RSU Cloud resource management selects configurations that improve the infrastructure delay significantly, with optimal number of service hosts. Over time and in face of dynamic loads, the configurations are selected as part of a Pareto Optimal Frontier, of non-dominated solutions. Any Pareto Optimal configuration is a candidate that can optimally, with respect to infrastructure delay and number of service hosts, minimize VM migrations, followed by control plane overhead. Our future work includes minimizing control plane modifications with an improved CRM heuristic and experimentation on GENI [24] testbed.

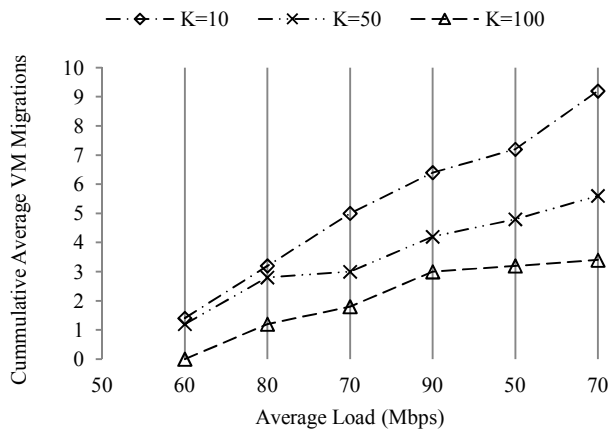

Fig. 7. Higher number of replications K, yields better results.

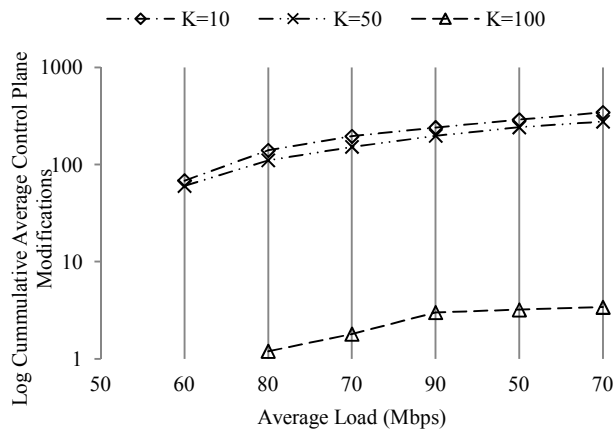

Fig. 8. Number of replications directly affects control plane modifications.